\documentclass[review]{elsarticle}

\usepackage{lineno,hyperref}
\journal{Journal of \LaTeX\ Templates}

\usepackage{graphicx}
\usepackage{subfigure}
\usepackage{amsmath}
\usepackage{epstopdf, epsfig}
\def\be{\begin{equation}}
\def\ee{\end{equation}}
\def\ba{\begin{array}}
\def\ea{\end{array}}
\def\bea{\begin{eqnarray}}
\def\eea{\end{eqnarray}}
\def\no{\nonumber}
\DeclareMathOperator{\arctanh}{arctanh}

\begin{document}

\begin{frontmatter}

\title{Phase modulated domain walls and dark solitons for surface gravity waves}

\author[mymainaddress,mysecondaryaddress]{Harneet Kaur}
%%\ead[url]{harneet.kaur13@gmail.com}

\author[mymainaddress]{Shailza Pathania}
%%\ead[url]{shalu.pathania7@gmail.com}

\author[mysecondaryaddress2]{Amit Goyal\corref{mycorrespondingauthor}}
\cortext[mycorrespondingauthor]{Corresponding author}
\ead{amit.goyal@ggdsd.ac.in}

\author[mymainaddress]{C. N. Kumar}
\ead{cnkumar@pu.ac.in}

\author[mysecondaryaddress3]{Daniela Milovic}
\ead{daniela.milovic@elfak.ni.ac.rs}

%%\cortext[mycorrespondingauthor]{Corresponding author}
%%\ead{cnkumar@pu.ac.in}

\address[mymainaddress]{Department of Physics, Panjab University, Chandigarh 160014, India}
\address[mysecondaryaddress]{Department of Physics, Government College for Women, Karnal 132001, India}
\address[mysecondaryaddress2]{Department of Physics, Goswami Ganesh Dutta Sanatan Dharma College, Chandigarh 160030, India}
\address[mysecondaryaddress3]{Faculty of Electronic Engineering, Department of Telecommunications, University of Nis, Aleksandra,
Medvedeva 14, 18000 Nis, Serbia}

\begin{abstract}
We report theoretical prediction of exact localized solutions for dynamics of surface gravity waves, at the critical point $kh\approx 1.363$, modelled by higher-order nonlinear Schr\"{o}dinger equation. The model possess domain walls (kink solitons) and dark solitons modulated through different phase profiles. The parametric domains are delineated for the existence of soliton solutions. The effect of wave parameters have been discussed on the amplitude of surface gravity waves. Our work is motivated by Tsitoura et al. \cite{prl}, on experimental and analytical observation of phase domain walls for deep water surface gravity waves modelled by nonlinear Schr\"{o}dinger equation.
\end{abstract}


\end{frontmatter}

%\linenumbers

\section{Introduction}
Domain walls (DWs) are the defects in the system which appear due to symmetry breaking. These are localized and shape preserving structures, which rise or descend from one asymptotic state to another. DWs appear as interface separating two stable states of the system such as  region of different alignment of spin in magnetic materials \cite{weiss}, regions of trivial and nontrivial phase for the phenomena of convection in pure and binary fluids \cite{cross1986, malo1990}, collision of travelling waves \cite{pio2001,malo1994,igor1995}. DWs connect the region having different amplitude and phase. These are also called as wave-jumps which appear as kink-type solutions of nonlinear evolution equations. DWs are also observed in nonlinear optics \cite{hael1994,malo} and surface gravity waves \cite{para1983,robert}. Recently, Tsitoura et al. \cite{prl} have reported experimental observation of phase domain walls for deep water surface gravity waves modelled by nonlinear Schr\"{o}dinger equation (NLSE).
\par In earlier works, stable wave pattern for deep water surface gravity waves
are studied numerically \cite{nick,cra}. Stability of these waves is analyzed by adding dissipation to the system \cite{segur2005}. Dissipation stabilizes the system by bounding the growth of perturbation before nonlinearity comes into play. Hammack et al. \cite{hammack2005} have experimentally studied progressive surface waves in deep water using standard and coupled NLSE. They analyzed the persistence of these waves in different parameter regime. In Ref. \cite{handerson2010}, authors have obtained
stable bi-periodic pattern on deep water in the presence of damping. In this work, we consider the surface gravity waves at the critical point, $kh\approx 1.363$ where $k$ is wave number and $h$ is undisturbed water depth, as it travels from deep to shallow water. Stability of periodic waves around this critical point is studied in the Refs. \cite{benja, feir}. Due
to phenomena of modulational instability, periodic wave train
becomes stable when $kh<1.363$ (shallow water) and unstable for $kh>1.363$ (deep water). Thus
for $kh>$1.363, modulation instability leads to formation of
solitons. At the critical value of $kh\approx 1.363$, cubic term vanishes
therefore higher order terms should be added to NLSE \cite{johnson}. In ref. \cite{johnson},
author have studied the dynamics of water waves near critical point
and presented the modulational instability regions . Slunyav extended this work and derived
higher-order NLSE for the case of vanishing cubic nonlinearity \cite{Slunyav}.
Grimshaw et al. \cite{grimshaw} obtained rescaled higher-order NLSE for water waves at
$kh\approx$ 1.363 and analyzed the behavior of wave as it propagates
from deep to shallow water. Similar kind of equations are studied in various physical contexts to obtain bright and dark solitons \cite{sed,kengne,ganda,liu,tsit}.
   \par In this paper, we analyzed higher-order NLSE to obtain kink and rational dark soliton
solutions, modulated through different phase profiles, for surface gravity waves at the critical point $kh\approx 1.363$ . The rational solution was firstly studied for $\phi^6$ model \cite{christ} and also explored in the field of nonlinear optics \cite{alka,meza1,meza2}. Here, our exact analysis takes recourse to identify these solutions for the ensuing model under certain parametric conditions. We demonstrate the control of these solutions for judicious choice of model and wave parameters.
\par An outline of the paper is as follow. In section 2,  model equation
and rescaled parameters are briefly discussed. In section 3, we
outline the procedure to obtain exact localized solutions. We present domain wall and dark solitons for the governing model.
We make the concluding remarks in section 4.

\section{Model Equation}
\par We considered the higher-order NLSE for the evolution of surface gravity waves at $kh\approx 1.363$ as
\cite{grimshaw}

\be \label{e1}
    i A_\tau+ \lambda_0 A_{\chi\chi}+\mu_0 |A|^2 A+i\alpha_0 |A|^2 A_\chi+i\beta_0 A^2 A^*_\chi+\nu_0 |A|^4
    A=0,\ee
where A is slowing varying complex envelope of carrier wave,
$\tau$ denotes time, $\chi$ is spatial coordinate, $\lambda_0, \mu_0,
\alpha_0, \beta_0$ and $\nu_0$ are model parameters. For surface
gravity waves near $kh\approx 1.363$, these parameters have the
values $\lambda_0=-0.2657 \omega k^{-2}, \mu_0=0.0002 \omega k^2,
\alpha_0=0.6833 \omega k, \beta_0=-0.2678 \omega k$ and
$\nu_0=-0.3864 \omega k^4$ \cite{Slunyav,grimshaw}. Introducing dimensionless variables $U=kA$, $x=k\chi$ and $\tau^{'}=\omega\tau$, the model equation (\ref{e1}) can be rewritten in dimensionless form as
    \be \label{e01}
    i U_{\tau^{'}}+ \lambda^{'} U_{xx}+\mu^{'} |U|^2 U+i\alpha^{'} |U|^2 U_x + i\beta^{'} U^2 U^*_x+\nu^{'} |U|^4 U=0,
    \ee
where $\lambda^{'}=-0.2657, \mu^{'}=0.0002,
\alpha^{'}=0.6833, \beta^{'}=-0.2678$ and
$\nu^{'}=-0.3864$. The number of
model parameters is reduced by dividing the Eq.
(\ref{e01}) with $\lambda^{'}$ and using the transformation $t=\lambda^{'} \tau^{'}$, which modifies the model parameters
as $\mu=\mu^{'}/\lambda^{'}$,
$\alpha=\alpha^{'}/\lambda^{'}$,
$\beta=\beta^{'}/\lambda^{'}$ and
$\nu=\nu^{'}/\lambda^{'}$. Rescaled
higher-order NLSE reads
    \be\label{e2}i U_t+ U_{xx}+\mu |U|^2 U+i\alpha |U|^2 U_x+i\beta U^2 U^*_x+\nu |U|^4 U=0,\ee
where the dimensionless parameters have the values $\mu=-0.0008,\alpha=-2.5717, \beta=1.0079$ and $\nu=1.4543$. Here, the dimensionless variables are related to model variables through the relations $A=U/k$, $\chi=x/k$ and $\tau=t/(\lambda^{'}\omega)$. For negative range of dimensionless variable $t$, the time variable $\tau$ will have positive range.

\section{Exact localized solutions}
  In order to solve the Eq. (\ref{e2}) for exact localized solutions, we choose ansatz of the form
  \be\label{e3}
    U(x,t)=a(\xi)e^{i(\psi(\xi)-\Omega t)},
  \ee
where $a$ and $\psi$ are real functions of $\xi$, $\xi=x-vt$ is a
travelling coordinate, and $\Omega,v$ are real parameters. Substituting
Eq. (\ref{e3}) into Eq. (\ref{e2}) and separating the
real and imaginary parts, we obtain \be\label{r1}
    a \Omega + av\psi^{'}+a^{''}-\psi^{'2} a+\mu a^{3}-\alpha
   a^{3} \psi^{'}+\nu a^{5}+\beta a^{3} \psi^{'}=0,
  \ee and
 \be\label{im1}
    -a^{'}v + 2 a^{'} \psi^{'}+\psi^{''} a+\alpha a^{'} a^2+\beta a^{'}
    a^2=0.
  \ee
Integrating Eq. (\ref{im1}) w.r.t. $\xi$, gives \be \label{new}
    \psi^{'}=\frac{I}{2 a^{2}}+\frac{v}{2}-\frac{(\alpha +\beta)a^{2}}{4},
\ee where $I$ is constant of integration. Substituting expression for $\psi^{'}$ into  Eq. (\ref{r1}), we obtain \be \label{master} a^{''}+b_1 a^5+b_2
a^3+b_3 a=\frac{I^2}{4 a^3},\ee
 where \be \no b_1=\nu-\frac{(\alpha+\beta)^2}{16}-\frac{\beta^2-\alpha^2}{4},~~~
b_2=\mu+(\beta-\alpha)\frac{v}{2},~~~ b_3=\Omega + \frac{v^2}{4}+\frac{(3\beta -\alpha)}{4}I.\ee
Eq. (\ref{new}) gives the phase profile for specific
form of $a(\xi)$ obtained by solving the Eq.
(\ref{master}). Eq. (\ref{master}) is studied in different contexts \cite{vyas,kumar}, and possesses a class of solutions such as rational solution \cite{christ}, bright and dark soliton \cite{behra}, and kink and periodic solution \cite{magyari}, in different parameter regimes. In this work, we present kink and rational solutions for which surface gravity waves have domain wall and dark soliton-like amplitude profiles, respectively.

\subsection{Domain walls}
In order to obtain kink type domain walls for Eq. (\ref{master}), we use transformation
$a=\sqrt{\sigma+\gamma}$, where $\gamma$ is a real constant. Using this transformation, Eq. (\ref{master}) can be rewritten as
% we obtain \be \label{e17}
%\frac{\sigma^{''}}{2}+2 b_3(\sigma+\gamma)+\frac{3}{2}b_2(\sigma+\gamma)^2+\frac{4}{3}b_1(\sigma+\gamma)^3=2c,\ee where c is constant of integration.
 \be \label{e18} \sigma^{''} +\frac{8}{3}b_1\sigma ^3 + 2(\frac{3}{2}b_2 + 4b_1 \gamma)\sigma^2 +
  2(2 b_3 +3b_2 \gamma + 4b_1\gamma^2)\sigma+4b_3\gamma + 3 b_2 \gamma^2 + \frac{8}{3} b_1\gamma^3-4k=0, \ee
  where $k$ is a constant of integration.
  For  $\gamma =\frac{-3 b_2}{8 b_1} $ and $ k = \frac{3}{128}\left(\frac{3 b_2^{3}-16 b_1 b_2 b_3}{b_1^{2}}\right)$,
  Eq. (\ref{e18}) reduces to the cubic elliptic equation of the form
\be \label{e19} \sigma^{''} + (4b_3 - \frac{9b_2^2}{8 b_1})\sigma + \frac{8}{3}b_2 \sigma^3 = 0.\ee
  For $b_2 < 0$ and $ (4b_3 - \frac{9b_2^2}{8 b_1}) > 0$, Eq. (\ref{e19}) has standard
 kink solution. The corresponding solution for Eq. (\ref{master}) reads  \be \label{e20} a(\xi) =
\sqrt{m~\mbox{tanh}(l\xi) +\gamma}, \ee such that $
m=\sqrt{\frac{27 b_2}{64 b_1}-\frac{3 b_3}{2b_2}} $ and $l =
\sqrt{2b_3-\frac{9 b_2^2}{16 b_1}}$. Parametric condition $b_2<0$
fixes the range of velocity as $v<\frac{-2 \mu}{\beta-\alpha}$ and
$ (4b_3 - \frac{9b_2^2}{8 b_1}) > 0$ gives range of integration
constant as $I>\frac{4}{(3 \beta-\alpha)}\left(\frac{9 b_2^{2}}{32
b_1}-(\Omega+\frac{v^2}{4})\right)$. The corresponding phase profile is given as
\begin{align}
    \no \psi(\xi)=\frac{v\xi}{2}&-\frac{1}{4} (\alpha +\beta)\left(\gamma\xi+\frac{m~\text{log}(\text{cosh}(l\xi))}{l}\right)\\
         \label{phase-eq3} &+\frac{I(-l\gamma\xi+m~\text{log}(\gamma~\text{cosh}(l\xi)+m~\text{sinh}(l\xi)))}{2\left(lm^2-l\gamma^2\right)} .
\end{align}
\begin{figure}[h!]
\begin{center}
\includegraphics[scale=0.65]{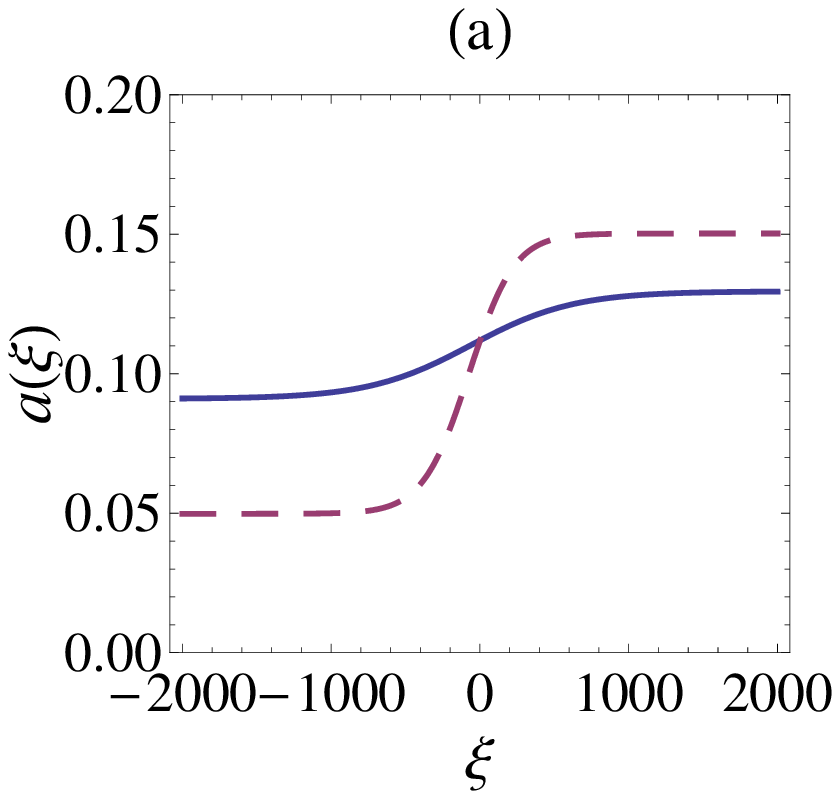}~~~~\includegraphics[scale=0.65]{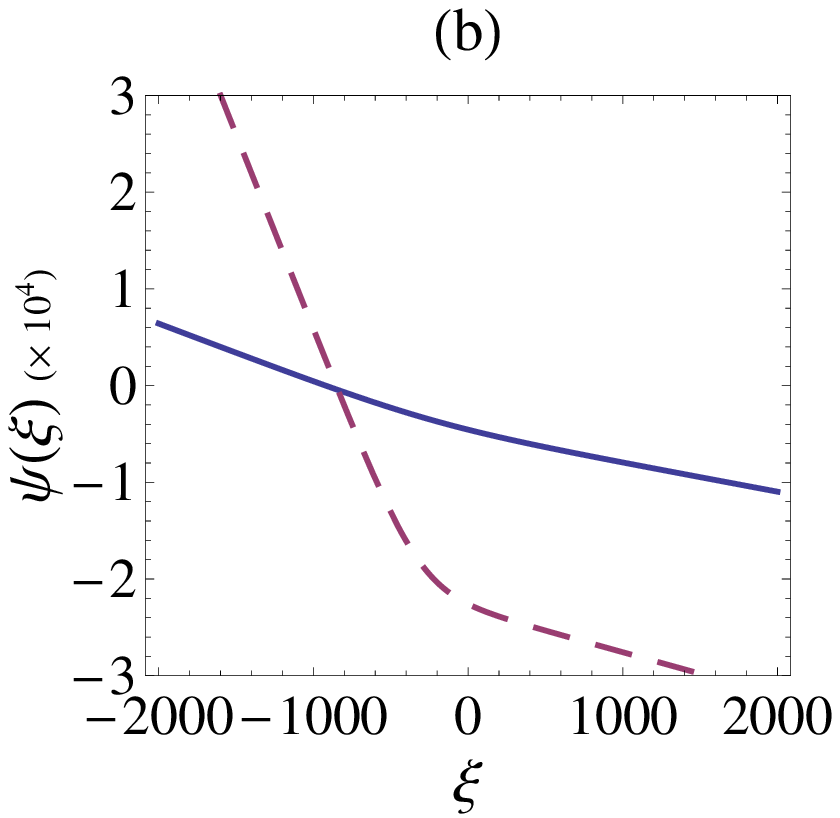}
\caption{\label{dkink4} (a) Kink solution
represented by Eq. (\ref{e20}), and (b) the corresponding phase profile, for $I=-0.1$ (solid line) and
$I=-0.2$ (dashed line). The other parameters are given in the text.}
\end{center}
\end{figure}
For the given values of model parameters (listed in the section 2),
the kink soliton given by Eq. (\ref{e20}) and phase profile are shown in Fig. \ref{dkink4} for different values of constant of
integration $`I$' and $v=-0.05$. For $I=-0.1$ and $-0.2$, the parameter $\omega$ is chosen as $0.14011$ and $0.28000$, respectively. For this choice, the other wave parameters are listed in Table \ref{table:1}.
\begin{table}[h!]
\centering
\begin{tabular}{ |c|c|c|c|}
 \hline
 I  & m & l & $\gamma$ \\
(constant of integration) & (amplitude) & (width) & (background) \\
\hline
 -0.1 & 0.00425 & 0.00147 & 0.01253 \\
\hline
-0.2 & 0.01005 & 0.00348 & 0.01253 \\
 \hline
\end{tabular}
\caption{Wave parameters for different values of constant of integration}
\label{table:1}
\end{table}
From Eq. (\ref{phase-eq3}), one can observe
that phase would be singular under the condition $m^2-\gamma^2=0$,
which imposes
constraint on the value of $I$ for specific choice of $v$ and
$\Omega$. The complex solution for Eq.
(\ref{e2}) reads
    \be \label{sol3}U(x,t)=\sqrt{m~\mbox{tanh}(l\xi) +\gamma}~e^{i(\psi(\xi)-\Omega t)}.\ee
\begin{figure}[h!]
\begin{center}
\includegraphics[scale=0.58]{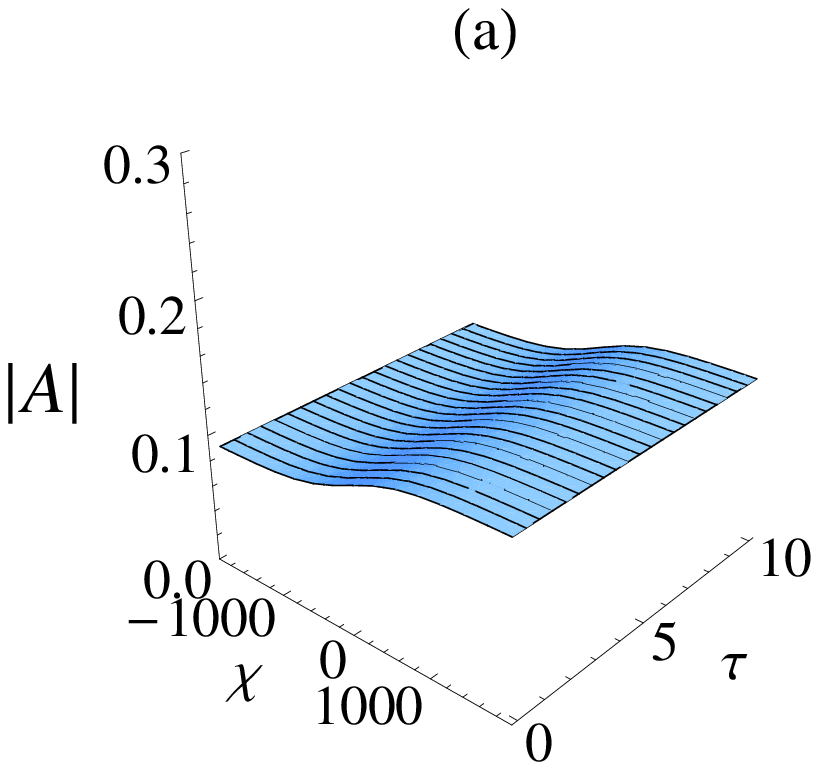}~~~~\includegraphics[scale=0.58]{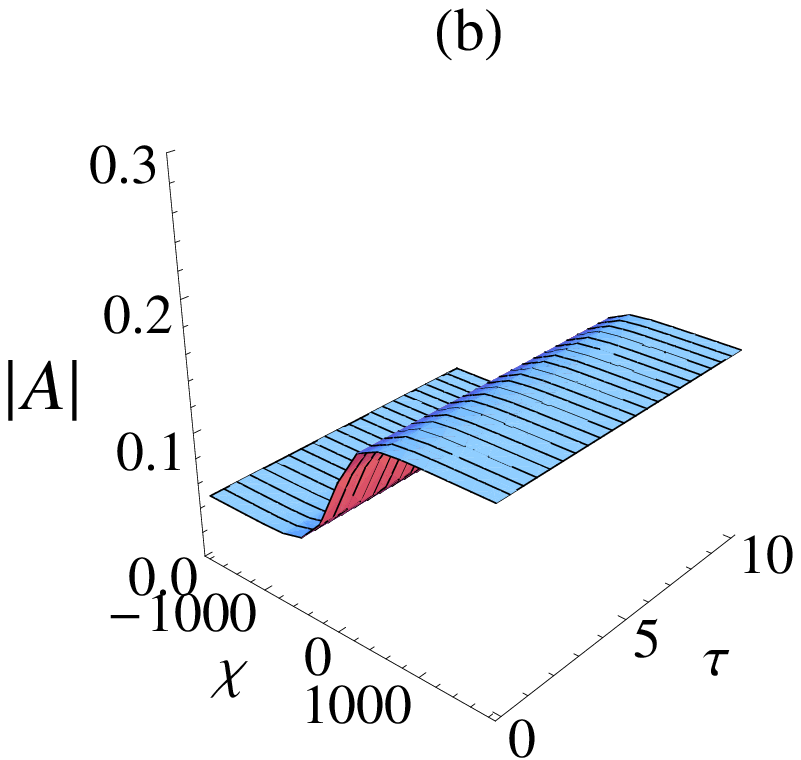}
\caption{\label{amp4}Amplitude profile of kink solitons for surface gravity waves modelled by Eq. (\ref{e1}) for different values of $I$, (a) $I=-0.1$ and (b) $I=-0.2$, respectively. The other parameters are given in the text. The variables $A$ and $\chi$ are plotted in units of $1/k$, and $\tau$ is in units of $1/\omega$.}
\end{center}
\end{figure}
In Fig. \ref{amp4}, we depicted the amplitude profile of surface gravity waves for the model equation (\ref{e1}), for different values of $I$, modulated through different phase profiles. In the limiting case of vanishing higher order contribution i.e. $\nu=0$, the allowed range of wave parameter $v$ or $\Omega$ will get modified. It is observed that for $\Omega = 0.28$ and $I = -0.2$, the allowed range of $v$ is $-0.0506 < v < 0.0004$ for quintic parameter $\nu = 1.4543$ and $-0.0213 < v < 0.0004$ for quintic parameter $\nu= 0$.
\subsection{Dark solitons} In this section, we demonstrate the existence of dark solitons for surface gravity waves. For $I=0$, Eq. (\ref{master})  admits rational solution of the
form \cite{christ,alka,amit} \be \label{edk1}
a(\xi)=\frac{p~\sinh(q \xi)}{\sqrt{\epsilon +\sinh(q \xi)^2 }},\ee
where $p=\sqrt{-\frac{2 b_3}{b_2}\left(\frac{2
\epsilon-3}{\epsilon-3}\right)}$,
$q=\sqrt{-b_3\left(\frac{\epsilon}{\epsilon-3}\right)}$ and
$\frac{b_1 b_3}{b_2^2}=\frac{3}{4}
\frac{(\epsilon-1)(\epsilon-3)}{(2\epsilon-3)^2}$.
The phase profile is given as \be\label{phase-eq1} \psi(\xi)=
\frac{v~\xi }{2}-\left(\frac{\alpha +\beta }{4}\right) p^2
\left(\xi -\frac{\sqrt{\epsilon} \arctanh(\frac{\sqrt{-1+\epsilon}
~\tanh(q \xi )}{\sqrt{\epsilon}})}{q \sqrt{-1+\epsilon }}\right ).
\ee
The corresponding complex envelope for the model Eq. (\ref{e2})
reads
    \be \label{sol1}U(x,t)=\frac{p~\sinh(q \xi)}{\sqrt{\epsilon +\sinh(q \xi)^2 }}~e^{i(\psi(\xi)-\Omega t)}.\ee
\begin{table}[h!]
\centering
\begin{tabular}{|c | c |c |}
 \hline
 \textbf{Range for $\epsilon$}     & \textbf{Condition on the coefficients $b_1$, $b_2$ and $b_3$}  & \textbf{Parametric constraints} \\
 \hline
 $\epsilon<0$             &   $b_1<0$, $b_2>0$ and $b_3<0$                        & $v>\frac{-2\mu}{\beta-\alpha}$,~$\Omega<-\frac{v^2}{4}$ \\
 \hline
 $0<\epsilon<1$           &   $b_1>0$, $b_2<0$ and $b_3>0$                        & $v<\frac{-2\mu}{\beta-\alpha}$,~$\Omega>-\frac{v^2}{4}$ \\
 \hline
 $1<\epsilon<\frac{3}{2}$ &   $b_1<0$, $b_2<0$ and $b_3>0$                        & $v<\frac{-2\mu}{\beta-\alpha}$,~$\Omega>-\frac{v^2}{4}$ \\
 \hline
 $\frac{3}{2}<\epsilon<3$ &   $b_1<0$, $b_2>0$ and $b_3>0$                        & $v>\frac{-2\mu}{\beta-\alpha}$,~$\Omega>-\frac{v^2}{4}$ \\
 \hline
 $\epsilon>3$             &   $b_1<0$, $b_2>0$ and $b_3<0$                        & $v>\frac{-2\mu}{\beta-\alpha}$,~$\Omega<-\frac{v^2}{4}$ \\
 \hline
\end{tabular}
\caption{Parametric constraint for different range of $\epsilon$}
\label{table:2}
\end{table}
 \begin{figure}[h!]
\begin{center}
\includegraphics[scale=0.65]{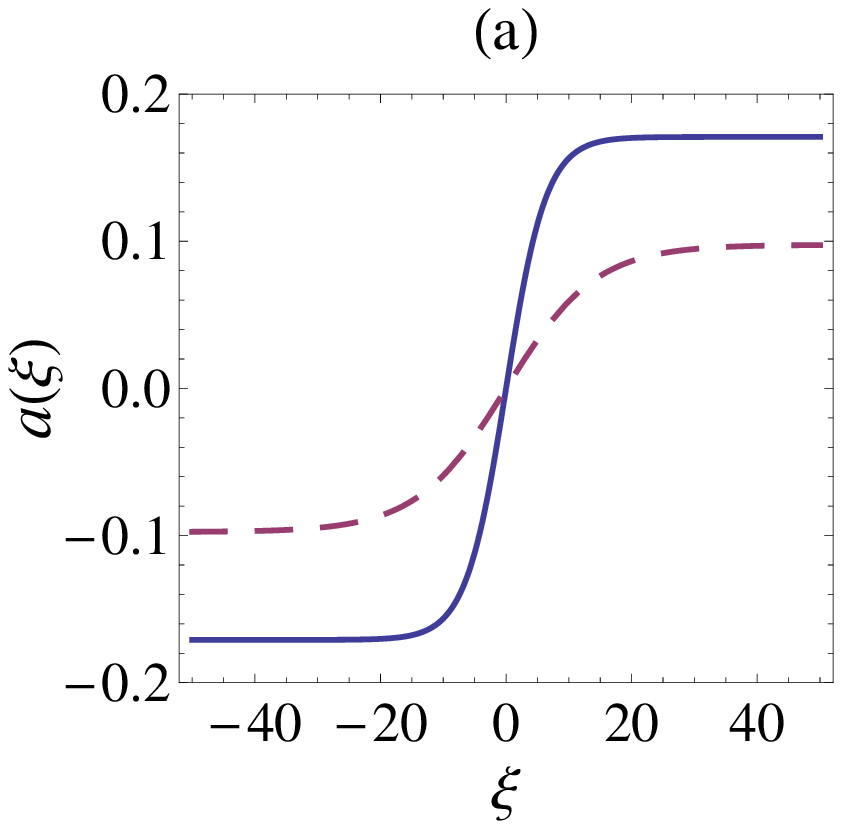}~~~~\includegraphics[scale=0.65]{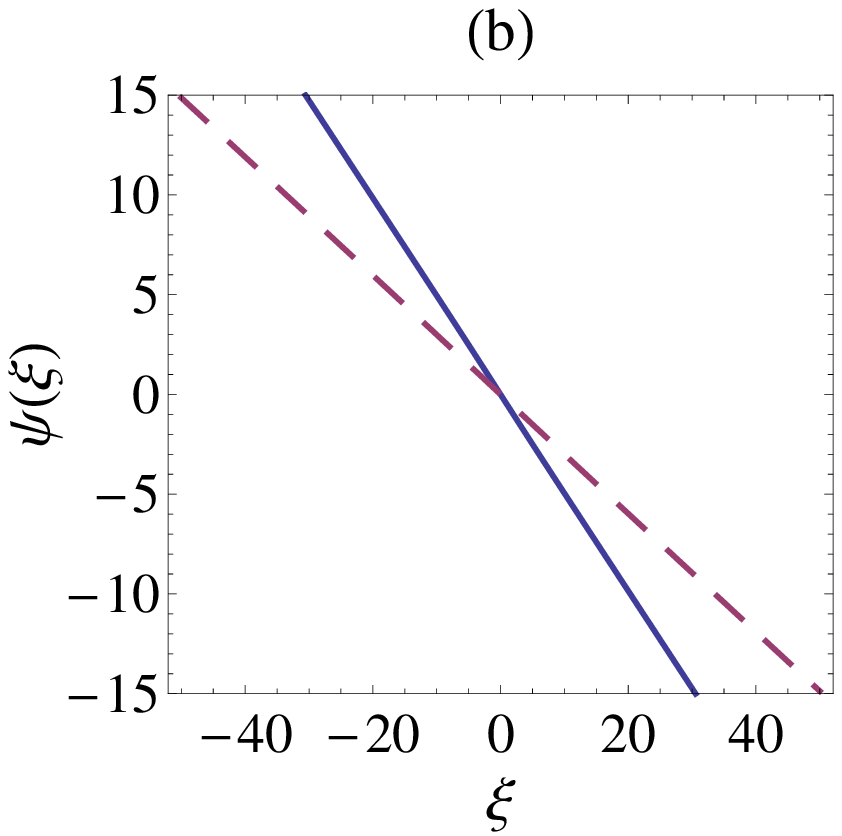}
\caption{\label{double1} (a) Kink solution
represented by Eq. (\ref{edk1}), and (b) the corresponding phase profile, for $v=-1.0$  (solid line) and $v=-0.6$ (dashed line). The other parameters are given in the text.}
\end{center}
\end{figure}
\begin{figure}[h!]
\begin{center}
\includegraphics[scale=0.58]{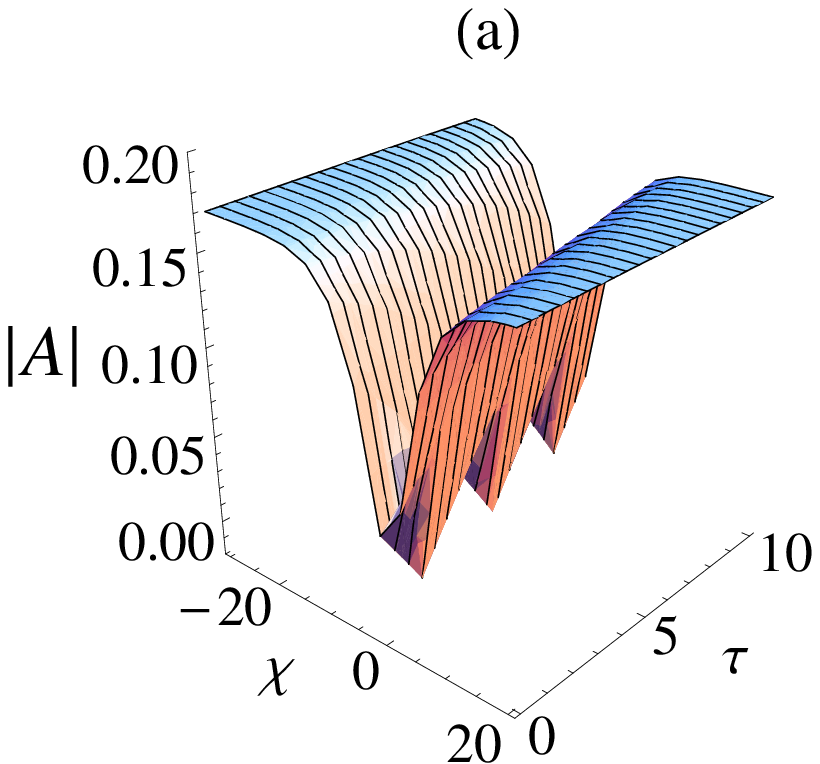}~~~~\includegraphics[scale=0.58]{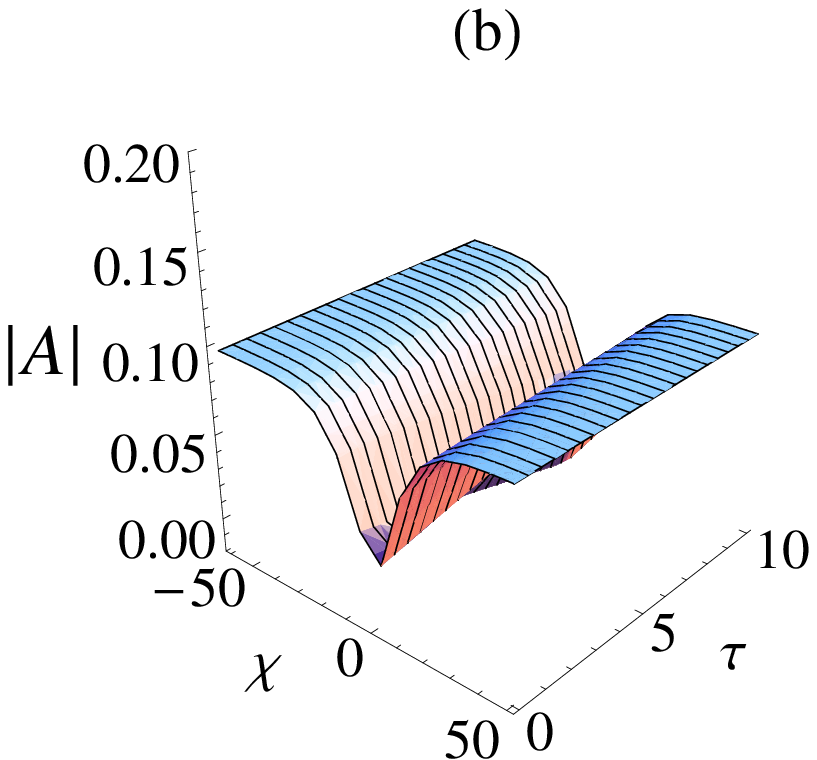}
\caption{\label{amp1}Amplitude profile of dark solitons for surface gravity waves modelled by Eq. (\ref{e1}) for different values of $v$, (a) $v=-1.0$ and (b) $v=-0.6$. The other parameters are given in the text. The variables $A$ and $\chi$ are plotted in units of $1/k$, and $\tau$ is in units of $1/\omega$.}
\end{center}
\end{figure}
Here, the parameters $p$ and $q$ should be real which impose the condition on the sign of coefficients
$b_1$, $b_2$ and $b_3$ for different ranges of $\epsilon$, under some constraints on wave parameters, as summarize in the table \ref{table:2}.
For the given set of model parameters, listed in the section 2,
the coefficient $b_1$ turns out to be greater than zero, hence
solution given by Eq. (\ref{edk1}) is possible in the range
$0<\epsilon<1$ (see table \ref{table:2}) and $v<0.0004$. The profile of kink
solution is shown in Fig. \ref{double1}(a) for different values of $v=-1.0$ $\&$ $-0.6$
and $\Omega=-0.2$ $\&$ $-0.08$, respectively. Here, it
is shown that the amplitude of the kink solution can be controlled by varying
the value of $v$ and $\Omega$ which in turn modulate the phase profile as shown in Fig. \ref{double1}(b). For the choice of
$v=-1.0$ and $-0.6$, the wave parameters comes out to be
$p=0.1709,~q=0.1544,~\epsilon=0.9687$ and
$p=0.0976,~q=0.0698,~\epsilon=0.9835$, respectively. The corresponding amplitude profile of the complex solution of Eq. (\ref{e1}) will be of the form of dark solitons as depicted in Fig. \ref{amp1}.

\par In the limit $\epsilon=1$, the
coefficients would be $b_2 < 0$, $b_3> 0$ and $b_1=0$ i.e. the
Eq. (\ref{master}) reduces to the cubic elliptic equation and the
solution given by Eq. (\ref{edk1}) reduces to standard kink
solution as
\be\label{e6} a(\xi) = a_0~\mbox{tanh}(m_0\xi),\ee
where $a_0= \sqrt{\frac{-b_3}{b_2}}$ and $ m_0 = \sqrt{\frac{b_3}{2}} $. Condition on $b_2$ and $b_3$ implies that $v<\frac{-2\mu}{\beta-\alpha}$ and $\Omega>-\frac{v^2}{4}$, and the condition $b_1=0$ imposes constraint on model parameters as $\nu=\frac{\beta^2-\alpha^2}{4}+\frac{(\alpha+\beta)^2}{16}$.

\section{Conclusions} We have examined the new class of exact localized solutions for surface gravity waves modulated through different phase profiles. We consider the model which governs water waves at critical value, $kh\approx 1.363$, and presented the domain wall and rational dark soliton for the set of model parameters obtained in the Refs. \cite{Slunyav,grimshaw} for fixed range of wave parameters $v$ and $\Omega$. We discuss the effect of wave parameters on the amplitude of surface gravity waves. Experimental realization of domain walls for deep water surface
gravity waves in the framework of NLSE, reported by Tsitoura et al. \cite{prl}, motivated us to look for domain wall solutions for surface gravity waves modelled by higher-order NLSE at critical point $kh=\approx 1.363$.

\section*{Acknowledgements}
A.G. and H.K. would like to thank Science and Engineering Research Board (SERB), Government of India for the award of SERB Start-Up Research Grant (Young Scientists) and student fellowship, respectively, under the sanction no: YSS/2015/001803, during the course of this work. S.P. would like to thank Department of Science and Technology (DST), Government of India for Junior Research Fellowship through Inspire Scheme under the sanction no: IF170725. We would like to thank the referees for their useful and constructive comments.

\bibliography{references}

\end{document}